\documentclass[prl,twocolumn,showpacs,preprintnumbers,amsmath,amssymb]{revtex4}
\usepackage[dvips]{graphics}

\begin{document}

\title{Quantumness without quantumness: \\entanglement as classical correlations in higher dimensions}

\author{Vlatko Vedral}
\email{ v.vedral@leeds.ac.uk} 
\affiliation{The School of Physics and Astronomy, University of Leeds,
Leeds LS2 9JT, England and \\
Quantum Information Technology Lab, Department of Physics,
National University of Singapore, Singapore 117542}

\date{\today}

\begin{abstract}
I exploit the formal equivalence between the ground state of a $d$ dimensional quantum system
and a $d+1$ dimensional classical Ising chain to represent quantum entanglement in terms of classical correlations
only. This offers a general ``local hidden variable model" for all quantum phenomena existing in one
dimension lower than the (hidden variable) classical model itself. The local hidden variable model is not contradicted by the implications of Bell's theorem. Formal theory is presented first and then exemplified by the quantum Ising
spin chain in a transverse magnetic field. Here I explicitely show how to derive any two site entanglement in the transverse model from the partition function of the classical Ising spin chain existing in two dimensions.  Some speculations are then presented regarding possible fundamental implications of these results.  \bigskip
\end{abstract}


\maketitle                           

Writing for special occasions is always a welcomed practice since it allows one to explore some more unusual
ideas and concepts that may not be fully appropriate for the standard mode of scientific communication. Some such speculative notions are perfectly suitable for festive publications and the author does not need to worry excessively about being carried away too far on his, as it were, wings of imagination.

Here I intend to take the full advantage of the mentioned ``author's" freedom that any festschrift offers and address a topic that I have been thinking about for quite some time on and off. I can only hope that Peter Knight, whom this issue is dedicated to, will not be too displeased with my contribution. With this apology in mind, I will now begin to expose my simple, yet - at least to me - very interesting, and hopefully somewhat original, idea.

In this paper I will explore if and how entanglement in many body systems can be represented only in terms of classical correlations. There has been a huge interest recently in entanglement in the condensed matter \cite{Vedral-rev} and various calculations have been performed linking entanglement to the notion of quantum phase transitions. In some sense, we can view entanglement as an order parameter, which has much the same properties as any other other order parameter such as magnetisation of a spin chain. While the exact physical function (if any?) of entanglement in many body systems still remains unclear, there is no doubt that entanglement can exist at various external condition (strengths of magnetic fields, temperatures, etc) even in the thermodynamical limit of a large number of subsystems. The existence of macroscopic entanglement has been confirmed experimentally and is very much beyond any doubt. Admittedly, its practical use is also not entirely unquestioned, but that need not concern us in this paper. 

It is well known that analogies exist between quantum and classical models in statistical physics \cite{Feynman,Yeomans,Justin}. The same formal analogies also exist between quantum dynamical and classical statistical models. More specifically, quantum propagators in $d$ dimensions are, in some sense, analogous to classical partition functions in $d+1$ dimensions. Likewise, the ground state (lowest) energy eigenvalue of the quantum system is the same as the free energy of the corresponding classical system. In general, the overall behaviour of a quantum system in some spatial dimension can be fully mapped into a classical model that is of one dimension higher. Furthermore, and very importantly for us here, short range quantum interactions are mapped onto short range classical interactions. Therefore, ``local" quantum systems lead to ``local" classical analogues.  

The $d\rightarrow d+1$ analogy can be best exemplified by a single quantum bit evolving under a simple Hamiltonian of the form:
\begin{equation}
H = E\sigma_z + \Delta \sigma_x
\end{equation}
where $\sigma_x, \sigma_z$ are the usual Pauli matrices. 
A single qubit is a zero dimensional system and, based on what I said, we should be able to represent it by a one dimensional classical system of some sort (a chain in other words). To find out what classical system it is, we need to re-write the evolution operator of the qubit as follows:
\begin{equation}
U = e^{-iHt} = e^{-iHt/m}e^{-iHt/m}...e^{-iHt/m}
\end{equation}
Inserting identities ($\sum_i |\sigma_i\rangle\langle\sigma_i|$, where $|\sigma_i\rangle$'s are eigenstates of $\sigma_z$) everywhere we obtain
\begin{equation}
\langle \sigma_m U |\sigma_1\rangle = \sum_{i_1,i_2,...i_m}\langle \sigma_{i_m}| e^{-iHt/m}|\sigma_{i_{m-1}}\rangle...\langle \sigma_{i_2}| e^{-iHt/m}|\sigma_{i_1}\rangle
\end{equation}
The most general transition element in this series can be written as \cite{Yeomans}
\begin{equation}
\langle \sigma_{j+1}| e^{-iHt/m}|\sigma_{j}\rangle = Ae^{h\sigma_j\sigma_{j+1} + K (\sigma_i + \sigma_{j+1})}
\end{equation}
where
\begin{eqnarray}
A & = & \sqrt{\frac{it\Delta}{m}}(1-\frac{(it)^2E^2}{m^2})^{1/4}\\
K & = & \frac{1}{4}\ln \frac{m^2-(it)^2E^2}{(it)^2\Delta^2}\\
h & = & \frac{1}{2}\ln \frac{m - it E}{m + it E}
\end{eqnarray}
Multiplying all the infinitesimal factors together we obtain
\begin{equation}
U = A^m \sum_{i_1,i_2,...i_m} e^{-H_I}
\end{equation}
where 
\begin{equation}
H_I = -2h\sum_i \sigma_i - K\sum_i \sigma_i\sigma_{i+1}
\end{equation}
is just the Hamiltonian for the (classical) Ising spin chain. The $\sigma$ elements are just numbers equal to $\pm 1$ (i.e. the eigenvalues of the $\sigma_z$ Pauli matrix). The propagator for a single qubit
(i.e. a zero dimensional two-level quantum system) is, therefore, seen to be the same as the partition function
of a one dimensional classical spin chain. This is not too surprising upon little reflection.
The qubit evolves from one instant to the next, which means that the state of the qubit at one instant ($j$) together with the Hamiltonian $H$ affects the state of qubit at the next instant ($j+1$). It is therefore expected that this dynamics is governed by the nearest neighbour interaction Ising Hamiltonian, where the neigbours are represented by the qubit values at two consecutive instants. 

An important observation for us is that the same holds true for systems in thermal equilibrium \cite{Suzuki}. Instead of the propagator, $U=e^{-iHt}$, the central object of study in thermodynamics is the partition function, $Z = tr e^{-\beta H}$ where 
$\beta = 1/kT$, $k$ being the Boltzmann constant and $T$ the temperature. Therefore, if we perform what is known as the analytic continuation, i.e. $it \rightarrow \beta$, all the results for computing the propagator immediately apply to the partition function. It is this fact that will be exploited in the remaining part of the article.  

I will now take a small detour to introduce some simple mathematics that will be useful for the remaining part of the paper. It will also indicate why the $d\rightarrow d+1$ correspondence holds in general, both for dynamical systems as well as systems in equilibrium. Suppose that we have a Hamiltonian with $p$ different parts $H = \sum_{j=1}^p A_j$. Then
\begin{equation}
e^{\sum_{j=1}^p A_j} = \lim_{n\rightarrow \infty} \{e^{A_1/n}...e^{A_p/n}\}^n
\end{equation} 
which is known as the Trotter formula \cite{Suzuki}. Inserting identities as before, we can write 
\begin{eqnarray}
\langle \alpha|e^{\sum_{j=1}^p A_p}|\alpha'\rangle & = & \lim_{n\rightarrow \infty} \sum_{\alpha_{ij}} \langle \alpha|  e^{A_1}|\alpha_{1,1}\rangle...\langle \alpha_{1,p-1}|  e^{A_p}|\alpha_{1,p}\rangle\nonumber \\
& \times & \langle \alpha_{1,p}| e^{A_1}|\alpha_{2,1}\rangle ...\langle \alpha_{2,p-1}| e^{A_p}|\alpha_{2,p}\rangle\nonumber \\
& \times & ... \langle \alpha_{n,p-1}| e^{A_p}|\alpha'\rangle
\end{eqnarray}
The double index of $\alpha$ now shows that a $d$ dimensional quantal model will be mapped onto $d+1$ dimensions classically. The above expansion is, in fact, also convenient as a calculational tool as long as the elements 
\begin{equation}
\langle \alpha|e^{A_j}|\alpha'\rangle
\end{equation}
are easy to compute (which will be the case in our examples below). The computational advantage lies in turning an exponential of non commuting matrices (difficult to calculate) into an exponential of numbers (admittedly more of them than the original matrices, but, nevertheless, much easier to calculate with). My point here will be that this calculational tool also has interesting conceptual implications in that I will use it to substitute entanglement by classical correlations. 

It is now at least intuitively clear that ground states of various quantum spin models will also be mapped in the same way into classical models with one dimension higher (as was first elaborated by Suzuki in \cite{Suzuki}). Since I want to study entanglement and how to represent it classically, I will concentrate on one dimensional quantum spin systems with nearest neighbour interactions. Their corresponding classical model will be a two dimensional Ising model, also with nearest neighbour interactions. This is a convenient and comforting fact previously mentioned: short range quantum models are always modeled by short range classical models. Everything I will say below is completely general and applies to any dimensions. 

One already extensively studied quantum spin chain is the Ising model in a transverse magnetic field \cite{Sachdev}. Since we know its entanglement properties at $T=0$ very well, I will focus on it as our main example. Its Hamiltonian is given by the following expression
\begin{equation}
H_{XZ} = - J \sum_{j=1}^M \sigma^z_j \sigma^z_{j+1} - B \sum_{j=1}^M \sigma^x_j
\end{equation}
where all the symbols have their usual meaning. Now, the partition function for this model, can be written as 
\begin{eqnarray}
Z & = & tr e^{\beta H_{XZ}} = tr e^{K\sum \sigma^z_{j}\sigma^z_{j+1} + \gamma \sigma^x_j} \nonumber \\
& = & \lim_{n \rightarrow \infty} A^{M}_n \sum_{\sigma = \pm 1} e^{\sum_{j=1}^M\sum_{k=1}^n K/n \sigma_{j,k}\sigma_{j+1,k} + K_n \sigma_{j,k}\sigma_{j,k+1}}   \nonumber
\end{eqnarray}
where
\begin{eqnarray}
A_n & = & \left(\frac{1}{2} \sinh \frac{2\gamma}{n}\right)^{n/2}\\
K_n & = & \frac{1}{2} \ln \coth \frac{\gamma}{n}
\end{eqnarray}
and 
\begin{eqnarray}
\gamma & = & \beta B \\
K & = & \beta J
\end{eqnarray}
As before, it is clear that the final form of the partition function just represents the classical two dimensional Ising model (as indicated by double indices in the partition function). 

All theremodynamical quantites in one dimension have their classical analogues in two dimensions. How about entanglement? Let us focus on one type of entanglement, that between nearest neighbouring spins in the one dimensional Ising chain. For that we need to know the following ground state averages: $\langle \sigma^x_i\rangle$, $\langle \sigma^x_{i+1} \rangle$, $\langle \sigma^x_i\otimes\sigma^x_{i+1}\rangle$, $\langle \sigma^y_i\otimes\sigma^y_{i+1}\rangle$ and $\langle \sigma^z_i\otimes\sigma^z_{i+1}\rangle$. All other averages are zero, which is not very difficult to show \cite{Sachdev}. Given that we know all the expectation values, we can reconstruct the two spin density matrix for nearest neighbours, from which can, in turn, infer when two nearest neighbouring spins are entangled (by performing the partial transposition, say). A details study of this was performed in \cite{Nielsen} and there is no need to repeat it here. Suffice it to say that entanglement certainly exists between nearest neighbours in the region where $J\le B$ (i.e. below what is known as the quantum critical region \cite{Sachdev}). 

How does the corresponding two dimensional Ising model simulate this entanglement? We will assume that the classical zero and one state are the eigenstate of the $z$ direction. We then do not have to worry about computing $\langle \sigma^z_i\otimes\sigma^z_{i+1}\rangle$, since this is the same classically or quantumly. But, we do need to figure out how to represent the other three averages listed above. Here the following identity helps: 
\begin{equation}
\langle \sigma| \sigma^x e^{\gamma/m \sigma^x}|\sigma'\rangle = e^{-2\gamma'_m \sigma\sigma'}  \langle \sigma| e^{\gamma/m \sigma^x}|\sigma'\rangle
\end{equation} 
where 
\begin{equation}
\gamma'_m = \frac{1}{2}\ln \{\coth(\gamma/m)\}
\end{equation}
Therefore, the correspondence between the one dimensional and two dimensional averages is: 
\begin{equation}
\sigma^x_j \rightarrow e^{-2\gamma'_m \sigma_{j,l}\sigma_{j,l+1}}
\end{equation}
Likewise,
\begin{equation}
\sigma^x_s \sigma^x_t\rightarrow e^{-2\gamma'_m \sigma_{s,l}\sigma_{s,l+1}}e^{-2\gamma'_m \sigma_{t,l}\sigma_{t,l+1}}
\end{equation}
Since the average of $z$ two point correlations is easy, so is in the $y$ direction since $\sigma^y = \sigma^x\sigma^z$. We are now in the position to represent all the quantum averages as some two dimensional classical Ising averages.  
And in this way we are automatically able to represent any genuine one dimensional quantum entanglement by a two dimensional Ising type classical correlation. Furthermore, these classical correlations are purely local. Namely, to simulate a quantum spin at a location $j$, we only need one more of its nearest neighbours (and nothing else) in the second dimension. 

It is clear that in general any type of multipartite quantum correlations can be represented in terms of the Ising classical correlations in higher dimensions. We only need to be able to find analogues of various spin correlation functions, but there is no doubt that such expressions do in principle exist (although, as above they may involve exponentials). 

How can quantum entanglement be fully mimicked by local classical correlations when we know that Bell's inequalities rule out any local hidden variable theories? It seems that we are somehow able to exploit the quantum to classical correspondence to construct a fully accurate hidden variable model for entanglement. This is not the case, however. Bell's inequalities stipulate that a quantum dichotomic variable (a variable with two outcomes) is to be simulated by a classical dichotomic variable. Here, on the other hand, we are using two classical Ising bits to simulate one quantum mechanical qubit. By allowing more capacity, classical spin chain can in this way reproduce all quantum correlations but of lower dimensionality. Note that a similar two-to-one ration exists in various quantum information protocols, such as super dense coding for example \cite{Vedral}. 

What does all this mean? Almost invariably the above analogy is used as a calculational tool. As we stressed, it is much easier to manipulate (exponentiate) classical spins, then it is to do so with the corresponding non-commuting (quantum) operators. This is exactly the same logic as in Feynman's path integral; computing propagators by exponentiating Hamiltonians is far more difficult than computing Feynman's integrals of exponentials of the corresponding Largangians. Various computational methods exploit this simple fact to solve complex solid state, or field theoretic problems. But, can there be more to the $d\rightarrow d+1$ correspondence? Is this quantum to classical correspondence trying to tell us something fundamental about the universe? 

I will now briefly try to outline two possible avenues for future research. The main issue I would like to discuss is whether this extra classical dimension that is used to simulate quantum properties is fictitious or if it can be made ``real" in some physical sense. For example, can entanglement in a three dimensional quantum universe lead to a classical four dimensional (classical) universe, where the fourth dimension is time or temperature or something else? Or, in the opposite direction, can a four dimensional classical universe lead to entanglement in three dimensions because our meansurement are always done in a short time interval? Can tracing out time in four dimensions lead to entanglement in three dimensions? A quantum state of a physical system always refers to a single time instance, but this may be how entanglement arises out of otherwise entanglementless universe in higher dimensions.    

The questions I am raising here are not entirely different to what others have already explored. Two other works are very much related to the raised questions. The first is by Page and Wootters \cite{Wootters}, on the so-called ``evolution without evolution" (this notion provided the main motivation for my title!). The second related work is by Sakharov \cite{Sakharov} on ``induced gravity". I now describe these two, if, for no other reason, then at least to contrast my idea with them. 

The work of Page and Wootters advances the idea that time can be (and is) encoded into correlations between systems and clocks (which are just some specially designated systems to measure temporal development - they are of no other special importance). The (timeless) state of the universe is according to Page and Wootters given by the following highly entangled state:
\begin{equation}
|\Psi_{universe}\rangle = |\psi^S_1\rangle |\psi^C_1\rangle + |\psi^S_2\rangle |\psi^C_2\rangle+...+|\psi^S_n\rangle |\psi^C_n\rangle
\end{equation}
where the first ket refers to the state of the system and the second to the state of the clock. 
This state does not evolve with time since it obeys the Hamiltonian constraint $H_{universe}|\Psi_{universe}\rangle=0$, but relative to the clock state, the state of the system does indeed change (as you can see by moving from one clock state to another, which then induces the usual Hamiltonian evolution in the system). Therefore, the temporal evolution in a static quantum state of the universe is encoded in the entanglement between different parts of the universe. A three dimensional spatial universe (plus entanglement) can in this way give rise to a four dimensional spacetime. It is clear that this idea is somehow related to the $d\rightarrow d+1$ correspondence I have been discussing here. But can it be turned upside down and made more in the spirit of the first part of the discussion? Before I discuss this issue let me first mention the second related idea. 

Sakharov's idea is somewhat different. It says that gravity, which is a manifestation of curvature of the four dimensional spacetime, is, in fact, not a fundamental force. It is just the result of vacuum fluctuations from the vacuum state arising from all other quantised fields in the universe (electromagnetic, weak and strong). Many have used this logic to argue that gravity therefore need not be quantised. Gravity is not fundamental and is just a quantum residual effect of some other more fundamental phenomena. But can Sakharov's induced gravity idea be generalised, and - at the same time - turned upside down? Namely, can a four dimensional curved, but classical, universe give rise to quantum mechanics in three dimensions? 

Suppose, for example, that we grant that the general relativistic spacetime manifold is given to us. This spacetime we assume to be fully classical, with one crucial difference: time is discretised and we can never measure properties at one single instant, but instead need to combine measurements from two neighbouring instants. This, according to our analysis would be manifested as entanglement in three dimensions. We therefore obtain quantum entanglement - the main characteristic train of quantum mechanics according to Schr\"odinger - from a fully classical universe!

As promised, I have now fully engaged in wild speculations. So, aside (and this is a big ``aside") from the fact that my idea has to be properly worked out theoretically \cite{Wonmin}, is there any hope that it can ever be tested experimentally? Page and Wootters's timeless universe looks more like a reinterpretation of quantum mechanics making it closer in spirit to General Relativity. It  seems difficult to see any difference between that and what we already know as quantum mechanics proper (or at least no one has pointed out any differences, as far as I can tell). Sakharov's thesis is more promissing, simply because it leads to higher order corrections in Einstein's field equations and these can, in principle, be tested. No such tests exist at present, however. The idea that entanglement is a classical correlation in four dimensions can likewise be tested if, for example, the extra dimension is indeed identified with time. The strength of entanglement in 3d would then determine the size of the corresponding temporal step. How divisible time is, is at least something that can be experimentally addressed, although perhaps we might in reality be very far from the scales at which this needs to be performed.

\textit{Acknowledgments}: The author acknowledges Janet Anders, Cristhian Avila, Libby Heaney, Jenny Hide, Wonmin Son and Mark Williamson 
for very stimulating discussions related to the nature of the ideas exposed in this paper. EPSRC is acknowledged for financial support.


\begin{thebibliography}{1}
\bibitem{Wonmin} This is now a work in progress in collaboration with Wonmin Son.
\end{thebibliography}
\end{document}